\documentclass[reqno]{amsart}
\usepackage{amscd,amssymb,url}
\usepackage[all]{xy}

\DeclareMathOperator{\tr}{tr}
\DeclareMathOperator{\Mod}{Mod}
\DeclareMathOperator{\Fred}{Fred}

\DeclareMathOperator{\Pro}{Pro}
\DeclareMathOperator{\Lin}{Lin}
\DeclareMathOperator{\Bun}{Bun}
\DeclareMathOperator{\ind}{ind}

\DeclareMathOperator{\Hom}{Hom}

\theoremstyle{plain}
\newtheorem{theorem}{Theorem}[section]
\newtheorem{lemma}[theorem]{Lemma}
\newtheorem{proposition}[theorem]{Proposition}

\theoremstyle{definition}
\newtheorem{definition}[theorem]{Definition}

\theoremstyle{remark}

\newtheorem{note}{Note}[section]

\numberwithin{equation}{section}
\numberwithin{figure}{section}

\newcommand{\cH}{{\mathcal H}}
\newcommand{\cS}{{\mathcal S}}
\newcommand{\cA}{{\mathcal A}}
\newcommand{\cE}{{\mathcal E}}
\newcommand{\cB}{{\mathcal B}}
\newcommand{\cK}{{\mathcal K}}
\newcommand{\cU}{{\mathcal U}}

\newcommand{\cP}{{\mathcal P}}
\newcommand{\cL}{{\mathcal L}}

\newcommand{\RR}{{\mathbb R}}
\newcommand{\ZZ}{{\mathbb Z}}

\newcommand{\C}{{\mathbb C}}
\newcommand{\R}{{\mathbb R}}
\newcommand{\Z}{{\mathbb Z}}

\newcommand{\redK}{\widetilde K} 

\newcommand{\UK}{{O_{\mathcal K}}}

\newcommand{\attn}[1]{\begin{center}\framebox{\begin{minipage}{9cm}#1\en 
d{minipage}}\end{center}}

\begin{document}

\title[Type I D-branes in an $H$-flux and twisted KO-theory]{Type I
D-branes in an $H$-flux and twisted KO-theory}

\author[V. Mathai]{Varghese Mathai}
\address[Varghese Mathai]
{Department of Pure Mathematics\\
University of Adelaide\\
Adelaide, SA 5005 \\
Australia}
\email{vmathai@maths.adelaide.edu.au }

\author[M. K. Murray]{Michael Murray}
\address[Michael Murray]
{Department of Pure Mathematics\\
University of Adelaide\\
Adelaide, SA 5005 \\
Australia}
\email{mmurray@maths.adelaide.edu.au }

\author[D. Stevenson]{Danny Stevenson}
\address[Danny Stevenson]
{Department of Pure Mathematics\\
University of Adelaide\\
Adelaide, SA 5005 \\
Australia}
\email{dstevens@maths.adelaide.edu.au }

\thanks{The authors acknowledge the support of the Australian
Research Council.}

\subjclass{81T30, 19K99}

\begin{abstract}
Witten has argued that charges of Type I $D$-branes
in the presence of an $H$-flux, take values in twisted $KO$-theory.
We begin with the study of real bundle gerbes and their holonomy.
We then introduce the notion of  real bundle gerbe $KO$-theory which we
establish is a geometric realization of twisted $KO$-theory.
We examine the relation with twisted $K$-theory, the Chern character and
  provide some examples.
We  conclude with some open problems.
\end{abstract}
\maketitle

\section{Introduction}

It has been argued by Witten \cite{Wit1}, that Type
I $D$-branes charges take values in $KO$-theory, based on explicit  
calculations by
\cite{blpssw} and \cite{SS}.  Now the Neveu-Schwarz $B$-field
   in Type I string theory is a local closed 2-form $B$ on $M$
which defines a class $H$ in $H^2(M, \ZZ_2)$, as argued in \cite{SS}  
and the paragraphs
after equation (5.7) in Witten (op.\! cit.\!).
When  the Neveu-Schwarz $B$-field $B$ is turned on, Type I  
$D$-branes
can be wrapped on a
submanifold $Z$ of spacetime only if the following equation holds,
\begin{equation}\label{anomaly}
[H |_Z] + w_2(Z) = 0 \quad \in H^2(Z, \ZZ_2),
\end{equation}
where $w_2(Z)$ denotes the second Stieffel-Whitney class of the tangent
bundle of $Z$.
Moreover, the normal bundle $NZ$ of $Z$ is $KO$-oriented if and only
if it has a spin structure,
i.e. $w_2(NZ) = 0$. However, this is not the case when there is a
Neveu-Schwarz $B$-field
if equation (\ref{anomaly}) is to be satisfied.
This and other considerations enabled Witten in (op.\! cit.\!) to deduce  
that
in the presence of a background flux $ H \in H^2(M, \ZZ_2)$, Type I  
$D$-branes
charges take values in {\em twisted} $KO$-theory, $KO(M, H)$.
This forces one to
work with twisted $KO$-theory.  In this note, we first of all give a  
geometric
interpretation of such 2-torsion $B$-fields as the holonomy of  
connections for real bundle
gerbes, which is the real analog of the bundle gerbe theory in  
\cite{Mur}
and which was briefly discussed in \cite{MS}. Just as the holonomy of  
an orthogonal connection
on a real line bundle represents the first Stieffel-Whitney class of  
the real line
bundle, we establish here that the holonomy of an orthogonal connection  
on a real bundle gerbe
represents  its  Dixmier-Douady class in $H^2(M, \ZZ_2)$,
which is sometimes referred to in
the physics literature as the t'Hooft class.
We also give a geometric realization of twisted $KO$-theory as the  
$KO$-theory
of real bundle gerbe modules, or briefly, real bundle gerbe $KO$-theory.
We give both finite dimensional and infinite
dimensional geometric realizations of twisted $KO$-theory, as both occur
naturally in examples.
This part can be viewed as the real analog of the
paper \cite{bcmms}.
Since dimension 10 is not crucial for our
analysis, it is possible that it applies to more general orientifolds in string
theory. As it was first shown Gukov's paper \cite{Gu}, $KO$ and $KSp$ 
groups of spheres classify
D-branes localized on orientifold planes of various dimension,
of which type I string theory is the simplest example. Therefore,
one could expect that the twisted $KO$-groups discussed in our paper
classify D-branes on orientifold planes in the presence of an H-flux,
so that the dimensionality of the orientifold plane is related to
the degree of suspension of the twisted $KO$-group. We thank
Sergei Gukov for these comments.

This paper is organised as follows. Section 2 summarises the theory of
real bundle gerbes. These are geometric objects that are associated with
degree 2, $\Z_2$-valued  \v{C}ech cohomology classes on $M$,
and is the real analog of the paper by Murray \cite{Mur}.
The notion of stable
equivalence of real bundle gerbes, which is essential for the  
understanding
of the sense in which the degree 2 class (known as the Dixmier-Douady
class of the real bundle gerbe) determines an associated bundle gerbe is
the subject of Section 3.  Section 4 studies the Chern-Weil
description of the Dixmier-Douady
class of real bundle gerbes, using a generalization of holonomy.
Here we encounter a problem that in a very small class of manifolds,
the holonomy does not determine the Dixmier-Douady
class of the real bundle gerbe, and we leave the geometric description
of the Dixmier-Douady
class of the real bundle gerbe for these manifolds
as an open problem in the final section.
Real bundle gerbe modules are introduced in section 5,
and the $KO$-theory of real bundle gerbe modules in section 6.
In Section 7, we establish the equivalence between the
real bundle gerbe $KO$-theory
and the twisted $KO$-theory as defined in \cite{DK} and \cite{Ros}
in terms of real Azumaya algebra bundles.
In section 8, an equivalent description of
  twisted $KO$-theory is given in  terms of infinite dimensional
real bundle gerbe modules with structure group $O_\cK$, as well as
other manifestations.
In Section 9, we briefly give the relation with the $KO$-theory of
continuous trace real $C^*$-algebras.
Sections 5-9 can be viewed as
the real analog of the paper by Bouwknegt, Carey, Mathai, Murray
and Stevenson \cite{bcmms}.
  In section 10, we define
the complexification map, which is a homomorphism from
$KO(M, H)$ to $K(M, \beta(H))$, where $\beta$ is the Bockstein
homomorphism. The complexification map complexifies the real bundle
gerbe into a complex bundle gerbe, as well as real bundle gerbe modules
into bundle gerbe modules. This enables us to define the twisted
Chern character from $KO$-theory to cohomology
and we establish that all the components of degree $(4n-2)$ vanish
for any positive integer $n$.
In this section, we also
consider examples where we write down non-trivial generators
for twisted $K$-theory, when the twist is a non-torsion class.
We end with some concluding remarks and open problems in section 11.

\section{real  bundle gerbes}
\label{bundle_gerbes}

\subsection{Real  bundle gerbes and Dixmier-Douady classes}

Here we define real  bundle gerbes in analogy with the definition given
in  \cite{Mur} in the complex case.  We refer to \cite{Mur,bcmms} for
standard notation regarding fibre products etc.
Recall that an orthogonal line bundle  $L \to M$ is a real line
bundle with a fibrewise
orthogonal inner product. For such a real line bundle the set of all
vectors of norm $1$ is
a principal $\ZZ_2$ bundle. Conversely if $P \to M$ is a principal
$\ZZ_2$ bundle then associated to it is a real
line bundle with fibrewise orthogonal inner product.  This is
formed in the standard way as the quotient of $P \times \RR$ by the
action of $\ZZ_2$ given by $(p, z)w = (pw, w^{-1}z)$ where $w \in  
\ZZ_2$.
The theory of real bundle gerbes  can use either principal $\ZZ_2$   
bundles
or equivalently orthogonal line bundles.
%
%
%
%
In the discussion below we will mostly adopt the latter point of view.
All maps between orthogonal line bundles will be assumed
to preserve the inner product unless we explicitly comment otherwise.

A real bundle gerbe
over $M$ is a pair $(L, Y)$ where
$\pi \colon Y \to M$ is a locally trivial fibration and
$L$ is an orthogonal line bundle $L \to Y^{[2]}$ on
the fibre product $Y^{[2]}$ which has a product, that is,
an isometric  isomorphism
\begin{equation}\label{gerbe}
L_{(y_1, y_2)} \otimes L_{(y_2, y_3)} \to L_{(y_1, y_3)}
\end{equation}
for every $(y_1, y_2)$ and $(y_2, y_3)$ in $Y^{[2]}$.
We require the product to be smooth in $y_1$, $y_2$ and
$y_3$ but in the interests of brevity we will not state the various
definitions needed to make this requirement precise, they  can be found  
in
\cite{Mur}.
The product is required to be
associative whenever triple products are defined. Also in \cite{Mur}
it is shown that the existence of the product and the associativity
imply isomorphisms $L_{(y, y)} \simeq \RR$ and $L_{(y_1, y_2)} \simeq
L_{(y_2, y_1)}^*$. We shall
often refer to a real bundle gerbe $(L, Y)$ as just $L$.
If the bundle gerbe $(L, Y)$ has $Y$ finite-dimensional
we will say that it is a \emph{finite-dimensional} bundle gerbe.
%
%
%

One defines such notions as morphisms of real bundle gerbes,
products and duals exactly as one does for the case of
complex bundle gerbes.  It is perhaps worth emphasising the
point that the product of two real bundle gerbes $(L,Y)$ and
$(J,X)$ is formed over the fibre product $Y\times_M X$ so that
$L\otimes J$ is the real line bundle over $(Y\times_M X)^{[2]}$
whose fibre at a point $((y_1,x_1),(y_2,x_2))$ is
$L_{(y_1,y_2)}\otimes J_{(x_1,x_2)}$.
If $J$ is a orthogonal line bundle over $Y$ then we can define a
real bundle gerbe $\delta(J)$ by $\delta(J) = {\pi_1^{-1}(J)}\otimes
\pi_2^{-1}(J)^*$, that is $\delta(J)_{(y_1, y_2)} = J_{y_2} \otimes  
J_{y_1}^*$,
where $\pi_i\,:\,Y^{[2]} \to Y$ is the map which omits the $i$th  
element.
The product on $\delta(J)$
is induced by the natural pairing
$$
J_{y_2}\otimes J_{y_1}^*\otimes J_{y_3}\otimes J_{y_2}^* \to
J_{y_3}\otimes J_{y_1}^*.$$
A real bundle gerbe which is isomorphic
to a  real bundle gerbe of the form $\delta(J)$ is  called {\em  
trivial}.
A choice of $J$ and  a real bundle gerbe
isomorphism $\delta(J) \simeq L$ is called
a {\em trivialisation}.  As is the case for
complex bundle gerbes, any two trivialisations
$J$ and $K$ of a real bundle gerbe $L$ differ by the
pullback of a real line bundle on $M$.
Hence the set of all trivialisations of
a given real bundle gerbe is naturally acted on by the set of all
orthogonal line bundles on $M$.  This is analogous to the way in which  
the
set of all trivialisations of a orthogonal line bundle $ L\to M$ is  
acted
on by $\ZZ_2$.

One can think of
real bundle gerbes as one stage in a hierarchy of objects with
each type of object having a characteristic class in $H^p(M, \ZZ_2)$.
For example if $p=1$ we have real line bundles on $M$, the  
characteristic
class is the first Stieffel-Whitney class.
When $p=2$ we have real bundle gerbes and
they have a characteristic class $d(L) = d(L, Y) \in H^2(M, \ZZ_2)$,
called the Dixmier-Douady class of $(L, Y)$, and is the
obstruction to the gerbe being trivial.  The class $d(L)$ is
constructed in essentially the same way as in \cite{Mur}.  Let
$P\to Y^{[2]}$ be the orthogonal
frame bundle associated to the orthogonal
line bundle $L$.  Choose an open cover $\{U_i\}_{i\in I}$ of $M$
such that there exist local sections $s_i\colon U_i\to Y$ of $\pi\colon
Y\to M$ and such that each non-empty intersection
$U_{i_1}\cap \cdots \cap U_{i_p}$ is contractible.  Form
maps $(s_i,s_j)\colon U_{ij}\to Y^{[2]}$ in the usual way
by sending $m\in U_{ij} = U_i\cap U_j$ to $(s_i(m),s_j(m))\in
Y^{[2]}$.  Let $P_{ij}$ denote the pullback bundle on $Y^{[2]}$.
The product on the bundle gerbe $L$ induces one on the principal
$\ZZ_2$ bundle $P$ and hence we have isomorphisms $P_{ij}\otimes
P_{jk}\to P_{ik}$.
%
%
%
%
%
%
Choose sections $\sigma_{ij}$ of $P_{ij}$ and
define $\epsilon_{ijk}\colon U_{ijk}\to \ZZ_2$ by
$\sigma_{ij}\sigma_{jk} = \sigma_{ik}\epsilon_{ijk}$.  The
associativity of the bundle gerbe product ensures that
$\epsilon_{ijk}$ satisfies the \v{C}ech $2$-cocycle condition
$\epsilon_{jkl}\epsilon_{ikl}^{-1}\epsilon_{ijl}\epsilon_{ijk}^{-1} =  
1$.

\begin{theorem}
\label{th:trivial}
A real bundle gerbe $(L, Y)$ has zero Dixmier-Douady class
precisely when it is trivial.
\end{theorem}

%
%

The proof is the obvious adaption of that in \cite{Mur}. If the
bundle is trivial we can make the
sections in the discussion above global and thus obtain a trivial
class.  If the
class is trivial then we can find $h_{ij}$ such that  
$(h_{ij}\sigma_{ij})
(h_{ik}\sigma_{ik})^{-1}(h_{jk}\sigma_{jk}) = 1$. We let $Y_i =
\pi^{-1}(U_i)$ and
define a bundle $Q_i$ by $(Q_i)_y = P_{(s_i(\pi(y)), y)}$. Over $Y_i
\cap Y_j$ the
bundle gerbe multiplication and the $h_{ij}\sigma_{ij}$ define
clutching isomorphisms $\hat\sigma_{ij} \colon
Y_i \to Y_j$ which satisfy $\sigma_{ij}\sigma_{ik}^{-1} \sigma_{jk}
=1$ and hence glue the $Q_i$ together
to define a global bundle $Q \to Y$.  It is straightforward to show
that this is a trivialisation
of $P$.


Notice that the same is true of the other objects in our hierarchy,
line bundles
are trivial if and only if their Stieffel-Whitney class vanishes.
The Dixmier-Douady class behaves as one would expect,
it is additive under forming products, it changes sign when
we take duals, and it is natural with respect to pullbacks.

\subsection{Lifting real bundle gerbes, Pfaffian line bundles and  
surjectivity}
\label{sec:lbg}
We will need one type of example of a real bundle gerbe in a number of  
places.
Consider a central extension of groups
$$
\ZZ_2 \to \hat G \to G.
$$
If $Y \to M$ is a principal $G$ bundle then it is well known that the
obstruction to lifting $Y$ to a $\hat G$ 
bundle $\hat{Y}$ that covers $Y$  is a  
class in $H^2(M, \ZZ_2)$.
It will be shown that a real bundle gerbe can be constructed
from $Y$, the so-called lifting real  bundle gerbe, whose Dixmier-Douady
class is the obstruction to lifting $Y$ to a $\hat G$ bundle. The  
construction
of the lifting real  bundle gerbe is quite simple. As $Y$ is a  
principal bundle
there is a map $g \colon Y^{[2]}\to G$ defined by $y_1 g(y_1, y_2) =  
y_2$.
We use this to pull back the $\ZZ_2$ bundle $\hat G \to G$ and form the
associated orthogonal
line bundle $L \to Y^{[2]}$. The real bundle gerbe product is induced
by the group
structure of $\hat G$.

Some finite dimensional examples of lifting real
bundle gerbes that we will be interested in
are those associated to the central extensions,
$$
\ZZ_2 \to SO(n) \to PO(n), \qquad \ZZ_2 \to Spin(n) \to SO(n).
$$
It is well known that the obstruction to lifting a principal $SO(n)$  
bundle
$Y$ to a $Spin(n) $ bundle is the second Stieffel-Whitney class   
$w_2(Y) \in H^2(M, \ZZ_2)$,
which is also equal to the  second Stieffel-Whitney class  of the  
associated orthogonal
vector bundle $Y\times_{SO(n)} \RR^n$.
In the physics literature, the real line bundle on $SO(n)$ that is  
associated to the
$Spin(n)$ double cover is known as the Pfaffian line bundle ${\rm  
Pfaff}$. It is occurs as a topological
  anomaly as follows. Real
vector bundles $E$ of rank $n$ over $S^1$ are determined by an
element $g \in SO(n)$ by the clutching
construction. So the self-adjoint family of
Dirac operators $\not\!\!\partial_{S\otimes E_g}$
on the real vector bundles $S\otimes E_g$, where
$S$ is the non-trivial real bundle of spinors on $S^1$,
is parametrized by the group $SO(n)$. Freed and Witten \cite{FW}
explained that the real determinant line bundle of this family,
which is essentially
defined as the highest exterior power of the kernel of the family,  
defines a non-trivial
real line bundle ${\rm Pfaff}$ on the group $SO(n)$,
which has the property that its lift
to $Spin(n)$ is the trivial real line bundle.
They also determine a connection and holonomy
of the Pfaffian line bundle.

Another interesting $\ZZ_2$ central extension that occurs in physics is
$$
\ZZ_2 \to Mp(n)\to Sp(n),
$$
where $Sp(n)$ denotes the symplectic group in $\RR^{2n}$ and
$Mp(n)$ the metaplectic group.
It is well known that the obstruction to lifting a principal $Sp(n)$  
bundle
$Y$ to a $Mp(n) $ bundle is the second Stieffel-Whitney class  $w_2(Y)  
\in H^2(M, \ZZ_2)$,
which is also equal to the  second Stieffel-Whitney class  of the  
associated symplectic
vector bundle $Y\times_{Sp(n)} \RR^{2n}$.

An infinite dimensional $\ZZ_2$ central extension that occurs in  
physics is
$$
\ZZ_2 \to SO(\cH)\to PO(\cH),
$$
for $\cH$ an infinite dimensional, separable, real Hilbert space. By  
Kuiper's
theorem, one knows that $SO(\cH)$ is a contractible group. Therefore  
$PO(\cH)$
is a model for the Eilenberg-Maclane space $K(\ZZ_2, 1)$, and so
$H^2(X, \ZZ_2) = [X, BPO(\cH)]$. Therefore,

\begin{theorem}
\label{th:PO}
The lifting real bundle gerbes of principal $PO(\cH)$ bundles
over $X$ are classified up to isomorphism
by their Dixmier-Douady invariant in $H^2(X, \ZZ_2)$. Moreover, every
element in $H^2(X, \ZZ_2)$ determines a lifting real bundle gerbe of a
principal $PO(\cH)$ bundle over $X$, up to isomorphism.
\end{theorem}

The following examples of real bundle gerbes follow the analogous
construction in \cite{Hit}.
Let $M$ be a compact Riemann surface, and choose a point $p\in M$.
Now cover $M$ with two open sets,
$U_1\cong \R^2$ a coordinate neighbourhood of $p$, and
  $U_0=M\setminus \{p\}$. Then
  $$U_0\cap U_1\cong \R^2\setminus \{0\}\cong S^1\times \R$$
  We define a real bundle gerbe ${\mathcal G}_p$ as follows.
Let $Y$ be the disjoint union of $U_0$ and $U_1$; then there is a 
submersion $Y\to M$.  
Consider the real line bundle $L=L_{01}$ on
$U_0\cap U_1$ which is the pull-back from $S^1$ of
the real line bundle whose first
Stieffel-Whitney class is the generator of
$H^1(S^1,\Z_2)$ and whose associated principal
bundle is the
non-trivial double cover of $S^1$. The choice of orientation on $M$  
gives an
orientation on $S^1$ and hence a choice of generator.
It is easy to see that $L\to Y^{[2]}$ defines a bundle gerbe.  
Since $M$ is compact, the Dixmier-Douady class of ${\mathcal G}_p$
is the generator of $H^2(M,\Z_2)\cong \Z_2$.
More generally, consider  an oriented codimension 2
submanifold $Q$ of a compact oriented manifold $M$. The
previous construction can be routinely generalized to this situation.
Take coordinate neighbourhoods
$U_\alpha$ of $M$ along $Q$, then
  $U_\alpha\cong (U_\alpha \cap Q)\times \R^2$
Now take
$U_0=X\backslash N(Q)$, where $N(Q)$ is the closure of a sufficiently  
small tubular
neighbourhood of $Q$ that is diffeomorphic to the disc bundle in the  
normal
bundle. Then $$U_0\cap U_\alpha \cong U_\alpha
\cap Q\times \{x\in \R^2:
\Vert x \Vert >\epsilon\}$$
and as before we
define $Y$ to be the disjoint union of
the $U_\alpha$ and $U_0$ so that we have a 
submersion $Y\to M$. 
Define the real bundle gerbe $L\to Y^{[2]}$ as follows.
Let $ L_{\alpha 0}$ be the pull-back by $x\mapsto x/\Vert x\Vert$ of the
real line bundle with non-trivial Stieffel-Whitney class on $S^1$.
The real line bundle $L_{\alpha\beta}= L_{\alpha 0}L^{-1}_{\beta 0}$ is  
defined on
$$U_{\alpha}\cap U_{\beta}\backslash N(Q) \cong (U_{\alpha}\cap  
U_{\beta}
\cap Q)\times \{x\in \R^2: \Vert x \Vert >\epsilon\}\simeq S^1$$
But  by construction $w_1(L_{\alpha\beta})=0$ is zero on $S^1$ and so
extends to a trivial real line bundle on the whole of $U_{\alpha}\cap  
U_{\beta}$.
This defines real line bundles on overlaps, and hence  the real line  
bundle $L\to Y^{[2]}$
It is not hard to see that the real bundle gerbe property is satisfied,
hence it defines a
real bundle gerbe ${\mathcal G}_Q$ on $M$ that is associated to $Q$.


\section{Stable isomorphism of real bundle gerbes}
There are many
real bundle gerbes which have the same Dixmier-Douady class but
which  are not isomorphic.
We can define a notion of {\em stable isomorphism} for real bundle
gerbes in exactly the same way as we can for complex bundle gerbes.
Just as is the case for complex bundle gerbes, two real
bundle gerbes have the same Dixmier-Douady class precisely
when there is a stable isomorphism between them.
\begin{definition}
\label{def:stableiso}
A stable isomorphism between real bundle gerbes  $(L, Y)$ and $(J, Z)$
is a trivialisation of $L^*\otimes J$.
\end{definition}
The proof of the following Proposition is entirely
analogous to the proof for complex bundle gerbes.
\begin{proposition}
\label{prop:stable}
A stable isomorphism exists from  $(L, Y)$ to $(J, Z)$
if and only if $d(L) = d(J)$.
\end{proposition}
If a stable isomorphism exists from $(L, Y)$ to $(J, Z)$
we say that $(L, Y)$ and $(J, Z)$ are stably isomorphic.
It follows easily that  stable
isomorphism is an equivalence relation.
Theorem \ref{th:PO} establishes that every class in $H^2(M, \ZZ_2)$ is
the Dixmier-Douady class of some real bundle gerbe. Hence we can deduce
from Proposition \ref{prop:stable} the
\begin{theorem}
\label{th:stableiso}
The Dixmier-Douady class defines a bijection between
stable isomorphism classes of  real bundle gerbes and $H^2(M, \ZZ_2)$.
\end{theorem}
Stable isomorphisms between real bundle gerbes can be composed
in exactly the same way as stable isomorphisms for complex
bundle gerbes.  We refer to \cite{bcmms} for more details.
Following Serre, Patterson \cite{Pat} established that every
element in $H^2(X, \ZZ_2)$ determines a principal $PO(n)$ bundle over  
$X$.
%
%
By the construction above this determines a lifting
bundle gerbe and  combining this with Theorem \ref{th:stableiso}, one  
has,

\begin{theorem}
\label{th:stableisofinite}
The Dixmier-Douady class defines a bijection between
stable isomorphism classes of  finite dimensional
real bundle gerbes and $H^2(M, \ZZ_2)$.
\end{theorem}
%
%
%
%

\section{Holonomy and Chern Weil theory of real Dixmier-Douady  
classes}\label{gerbe hol}
   A $U(1)$ bundle gerbe with
connection and curving determines a Deligne cohomology class.
Interpreted as a Cheeger-Simons differential character this is a pair  
consisting
of a homomorphism
$h \colon Z_2(M) \to U(1)$ and a three-form $\omega $ on $M$ satisfying
$$
h(\partial \sigma)= \exp\left( \int_\sigma \omega \right)
$$
for any three-cycle $\sigma$.
The map $h $ applied to smooth surfaces is just the holonomy and
$\omega $ is the three-curvature.  We will concentrate attention
for the remainder of this section on the case where
   $H_2(M, \ZZ)$ is generated by fundamental strings,
which are just smooth maps of
surfaces into $M$. $h$ will be the holonomy
  for a real $\ZZ_2$ bundle gerbe, which takes its values in $\ZZ_2$.


For a real $\ZZ_2$ bundle gerbe the connection is flat and its curvature is  
zero.
We can therefore take as a curving just the zero two-form on $Y$ and
it follows that the three-curvature on $M$ is zero.  In such a case
$h$ applied to boundaries is one so that it factors to a homomorphism
$h \colon H_2(M, \ZZ) \to \ZZ_2$.

On the other hand for a real $\ZZ_2$ bundle gerbe we have the
Dixmier-Douady class in $H^2(M, \ZZ_2)$ which we can map,
as in the universal coefficient theorem cf. \ref{uct}, to $\Hom(H_2(M,  
\ZZ), \ZZ_2)$.


We wish to show that this map is just the holonomy. As we are
assuming that $H_2(M, \ZZ)$ is generated by fundamental strings,
  it suffices to prove this for $M = \Sigma$ a smooth
surface.

Let $\Sigma$ then be a closed surface. Choose an open cover $U_0$, $U_1$
where $U_0$ is a neighbourhood of a point in $\Sigma$ and $U_1$ is a  
slightly
enlarged open neighbourhood of the complement of $U_0$ chosen so that
$U_0 \cap U_1$
is an annular region.


Consider a $\ZZ_2$ real bundle gerbe $(P, Y)$ over $\Sigma$.
This classified by a
class in  $H^2(\Sigma, \ZZ_2) = \ZZ_2$.
Let $Y_i = \pi^{-1}(U_i)$.
As $H^2(U_i, \ZZ_2)= 0$ the restriction
of $(P, Y)$  to  $U_0$ and $U_1$ is trivial. So we can find a
$\ZZ_2$ bundles $Q_i \to Y_i $ such that $\delta(Q_i) = P$ for
each $i=0, 1$. Denote by
$Q_{01}$ the $\ZZ_2$ bundle over $U_0 \cap U_1$ which
is the difference between $Q_0$ and  $Q_1$. If we put on $(P, Y)$ a
flat connection then this is
inherited by $Q_{01}$ and its holonomy is either $+1$ if $Q_{01}$
is trivial or $-1$ if $Q_{01}$ is not trivial.  The original bundle
gerbe $(P, Y)$ is stably isomorphic to its pullback to the
disjoint union of the two open sets $U_0$ and $U_1$ and hence is
trivial if and only if we can find $\ZZ_2$ bundles
$T_i \to U_i$ with $T_0 = Q_{01} \otimes T_1$ on $U_0 \cap U_1$.
As $U_0$ is a disk we have that $T_0 $ is trivial. Also from
   Mayer-Vietoris we have that $H^1(U_1, \ZZ_2) = H^1(\Sigma, \ZZ_2)$
   so that $T_1$ extends to all of $\Sigma$, hence restricted
   to $U_0 \cap U_1$ it is trivial. Thus $Q_{01}$ is trivial.
   We conclude that $(P,Y)$ is trivial
   if and only if $Q_{01}$ is trivial and hence the class in
$H^2(\Sigma, \ZZ_2) = \ZZ_2$
   is the same as the class determined by $Q_{01}$ in $H^1(U_0 \cap
U_1, \ZZ_2) = \ZZ_2$.

Now we calculate the holonomy.  Because $H^3(\Sigma, \ZZ) = 0$ we
can find a $U(1) $ bundle $R \to Y$ such that $\delta(R) =
   R \times_{\ZZ_2} U(1)$. Denote by $R_i$ the $U(1)$
bundle over $U_i$ which is the difference between the trivialisations
$Q_i \times_{\ZZ_2}U(1)$ and $R$.
We have
$\pi^{-1}(R_0) \otimes Q_0 = R = \pi^{-1}(R_1) \otimes Q_1 $ and
$Q_0 = \pi^{-1}(Q_{01}) \otimes Q_1$ and hence on $U_0 \cap U_1$ we  
must have
$$
R_1 = Q_{01} \otimes R_0.
$$
General bundle gerbe theory tells us that we can choose a connection
on $R$ which under $\delta$ is equal to the connection on $Q$. This  
means
that we can choose connections $D_i$ on $R_i$ such that on $U_0 \cap  
U_1$
we have
$$
D_1 = D + D_0
$$
where $D$ is the flat connection on $Q_{01}$. We claim that the  
holonomy of the
connection and curving of $(P, Y)$ is the holonomy of $D$ around any
non-trivial loop in $U_0 \cap U_1$.  To see this note first that $F_0$,
the curvature of $D_0$, is a closed two-form on a disk so we have
$F_0 = d\alpha_0$ and hence $D_0 -\alpha_0$ is flat. If we restrict  
$\alpha_0$
to $U_0 \cap U_1$ we can extend it to $\alpha_1 $ on $U_0 \cap U_1$.
Letting $\tilde D_i = D_i - \alpha_i$ we note that we still have
$$
\tilde D_1 = D + \tilde D_0
$$
but now $\tilde D_0$ is flat. The holonomy of the bundle gerbe is the
exponential of the integral of the global two-form $F$ which is
$F_1$ on $U_1$ and $F_0 $ on $U_0$ but $F_0 = 0$ so
the integral is just over $U_1$. Hence the holonomy
of the bundle gerbe is the holonomy of $D_1$ over $U_0 \cap U_1$
but that is the product of the holonomy of $D$ and the holonomy
of $D_0$. The latter is the identity as $F_0 = 0$.  So we have
the required result.

These results apply also to  lifting bundle gerbes which for $SO(3)$
bundles we can also see directly.
Let $P \to \Sigma$ be a principal $SO(3)$ bundle. This is classified
by the class
in $H^2(\Sigma, \ZZ_2) = \ZZ_2$ which is the obstruction to lifting
to $SU(2)$. Indeed if $P$ is trivial  it lifts to $SU(2)$ and the
class vanishes. If $P$ is not
trivial and it lifts to an $SU(2)$ bundle $\hat P$ then, as $SU(2)$
is simply-connected
$\hat P$ is trivial so that $P = \hat P / \ZZ_2$ is trivial. Hence $P$  
is
trivial if and only if the class in $H^2(\Sigma, \ZZ_2) $ vanishes.
This also implies that $P \to \Sigma$ is trivial on restriction to
$U_0$ and $U_1$
and $H^2(U_i, \ZZ_2) = 0$ so there is a
clutching map $U_0 \cap U_1 \to SO(3)$ and, as $\pi_1(SO(3)) = \ZZ_2$
there are two cases,
either the map is homotopic to a constant and $P$ is trivial or the map  
is
not homotopic to a constant and $P$ is not trivial. Equivalently the
clutching map
defines a class in $H^1(U_0 \cap U_0, \ZZ_2)  = \ZZ_2 $ which
vanishes precisely
when $P$ is trivial i.e as an element of $\ZZ_2$ it is the same as
the class in $H^2(\Sigma, \ZZ_2)$.

One can ask the question as to what extent does the holonomy of the  
real bundle gerbe
determine the bundle gerbe? This is revealed by the
universal coefficient theorem,
which in this case is the split exact sequence
\begin{equation}\label{uct}
0\to {\rm Ext}(H_1(M, \ZZ), \ZZ_2) \rightarrow H^2(\Sigma, \ZZ_2)
{\rightarrow} \Hom(H_2(M, \ZZ), \ZZ_2) \to 0
\end{equation}
It is well known that  ${\rm Ext}(H_1(M, \ZZ), \ZZ_2) = 0$ whenever
the torsion subgroup of $H_1(M, \ZZ)$ has only odd order torsion
subgroups,  cf. \cite{Hat}, page 195.

\section{Real bundle gerbe modules}

We can also define modules for real bundle gerbes in an
almost identical fashion to the complex case.
Thus let $(L, Y)$ be a real bundle gerbe
over a manifold $M$ and let $E \to Y$
be a finite rank, orthogonal vector bundle.
Assume that there is a vector bundle isomorphism
\begin{equation}
\label{eq:bgmod}
\phi \colon L\otimes \pi_1^{-1}E \stackrel{\sim}{\to}
\pi_2^{-1}E
\end{equation}
which is compatible with the real
bundle gerbe multiplication in the obvious sense.
In such a case we call $E$ a  real bundle gerbe module and say
that the real bundle gerbe acts on $E$.

We define two real bundle gerbe modules to be isomorphic if they are
isomorphic as real vector bundles and the isomorphism
preserves the action of the real bundle gerbe.
Denote by $\Mod_\R(L)$ the set of all isomorphism classes of
real bundle gerbe modules for $L$.
If $(L, Y)$ acts on $E$ and also on $F$ then it acts on $E\oplus F$
in the obvious diagonal manner.  The set  $\Mod_\R(L)$ is therefore
a semi-group.  Note that if $E$ is a real bundle gerbe module,
then it is of rank a multiple of $2^{[j/2]}$.

Suppose the real bundle gerbe $L$ has a trivialisation
$K$.  Then we have fibre-wise isomorphisms
$L_{(y_1, y_2)} = K_{y_2} \otimes
K_{y_1}^*$ for points $y_1$ and $y_2$
of $Y$ lying in the same fibre over $M$.
Therefore if $E$ is an $(L, Y)$ real bundle gerbe module
then we have fibre-wise isomorphisms $K_{y_2} \otimes
E_{y_2} \simeq
K_{y_1} \otimes E_{y_1}$.  In fact it can be
shown (see \cite{bcmms}) that the bundle $K\otimes E$
descends to a real vector bundle on $M$.
Conversely
if $F$ is a real vector bundle on $M$ then $L$ acts on $K\otimes  
\pi^{-1}(F)$. Denote
by $\Bun_\R(M)$ the semi-group of all isomorphism classes of real vector
bundles on $M$. Then we have.
we have
\begin{proposition}
\label{prop:trivialcase}
A trivialisation of  $(L, Y)$ defines  a semi-group
isomorphism from $\Mod_\R(L)$ to $\Bun_\R(M)$.
\end{proposition}

Notice that this isomorphism is not canonical but depends on the choice
of the trivialisation. If we change the trivialisation by tensoring
with the pull-back of a real line bundle $J$ on $M$ then the  
isomorphism changes
by composition with the endomorphism of $\Bun_\R(M)$ defined by  
tensoring
with $J$.

Recall that a stable isomorphism from a real bundle gerbe $(L, Y)$
to a real bundle gerbe $(J, X)$ is a trivialisation of $L^*\otimes J$.  
This
means there is a real line bundle $K \to Y \times_f X$ and an  
isomorphism
$ L^*\otimes J \to \delta(K)$ or, in other words, for
every $(y_1, y_2)$ and $(x_1, x_2)$ we have an isomorphism
$$
       L_{(y_1, y_2)}^*\otimes J_{(x_1, x_2)}
\to K_{(y_2, x_2)} \otimes K_{(y_1, x_1)}^*.
$$

Let $E \to Y$ be a real bundle gerbe module for $L$ and define
$\hat F_{(y, x)} = K_{(y,x)}^* \otimes
E_y$ a bundle on $Y \times_f X$.
We have isomorphisms
\begin{align*}
\hat F_{(y_2, x)} &= K_{(y_2,x)}^* \otimes E_{y_2}\\
                        &= K_{(y_1, x)}^* \otimes L_{(y_1, y_2)} \otimes
E_{y_2} \\
                        &= K_{(y_1, x)}^* \otimes E_{y_1} \\
                        &= \hat F_{(y_1, x)}.
\end{align*}
It is not hard to see that the bundle $\hat{F}$ must in
fact be the pullback under the map $Y\times_f X\to X$
of an orthogonal bundle $F$ on $X$.  The bundle $F$ on
$X$ is in fact a $(J,X)$ module since we have isomorphisms
\begin{align*}
J_{(x_1, x_2)} \otimes F_{x_2} &=J_{(x_1, x_2)} \otimes K_{(y,
x_2)}^*\otimes E_y \\
                           &= K_{(y, x_1)}^* \otimes E_y \\
                           & = F_{x_1}
\end{align*}
Therefore a choice of stable isomorphism defines a map
$$
\Mod_\R(L) \to \Mod_\R(J).
$$
In a similar fashion we can define an inverse map
$$
\Mod_\R(J) \to \Mod_\R(L).
$$
Hence we have the following real analogue
of Proposition 4.3 of \cite{bcmms}.
\begin{proposition}
\label{prop:module}
A stable isomorphism from $(L, Y)$ to
$(J, Y)$ induces an isomorphism of semi-groups
between $\Mod_\R(L)$ and $\Mod_\R(J)$.
\end{proposition}

Note that, as in the trivial case, this isomorphism is not
canonical but depends on the choice of stable isomorphism. Changing the
stable isomorphism by tensoring with the pull-back of a real line bundle
$J$ over $M$
changes the isomorphism in Prop.  \ref{prop:module}
by composition with the endomorphism
of $\Mod_\R(J)$ induced by tensoring with the pull-back of $J$.

There is a close relationship between real bundle
gerbe modules and bundles of real projective spaces. Recall that a
bundle of real projective spaces
$\cP \to M$ is a locally trivial fibration whose fibres
are isomorphic
to $P(V)$ for $V$ a real Hilbert space, either finite or infinite  
dimensional,
and whose structure group is $PO(V)$. This  means that there is a
$PO(V)$ bundle $X \to M$ and $\cP = X \times_{PO(V)} P(V)$.
Associated to $X$ is a
lifting real bundle gerbe $J \to X^{[2]}$ and a Dixmier-Douady class.
This Dixmier-Douady class is the obstruction to $\cP$ being the
projectivisation of a real vector bundle.
The lifting real bundle gerbe acts naturally
on the real bundle gerbe module $E = X \times H$ because each $J_{(x_1,
x_2)} \subset O(V)$ by
construction \ref{sec:lbg}.

Let $(L, Y)$ be a
real bundle gerbe and $E \to Y$ a real
bundle gerbe module. Then the projectivisation
of $E$ descends to a real projective
bundle $\cP_E \to M$ because of the
real bundle gerbe action. It is
straightforward to check that the class of this
real projective bundle is $d(L)$.
Conversely if $\cP \to M$ is a real projective
bundle with class $d(L)$ the associated
lifting real bundle gerbe has class
$d(L)$ and hence is stably isomorphic to
$(L, Y)$. So the module on which the
lifting real bundle gerbe acts defines a module
on which $(L, Y)$ acts.  From the
discussion before \ref{prop:module} one
can see that if  two modules
are related
by a stable isomorphism they give rise
to the same real projective bundle
on $M$.   We also  have that $E \to Y$
and $F \to Y$ give rise to
     isomorphic projective bundles on $M$
if and only if there is a line
bundle $K \to M$ with $E = \pi^{-1}(K)\otimes F$.
Denote by $\Lin_\R(M)$ the group
of all isomorphism classes of real  line bundles on $M$. Then this acts  
on
$\Mod_\R(L)$ by $E \mapsto \pi^{-1}(K) \otimes E$ for any real line
bundle $K \in \Lin_\R(M)$.  If $[H] \in H^2(M, \ZZ_2)$, denote
by $\Pro_\R(M, [H])$ the set of all isomorphism classes of
real projective bundles with class $[H]$.
Then we have

\begin{proposition}
\label{prop:proj}
If $(L, Y)$ is a real bundle gerbe then the  map
which associates to any element of $\Mod_\R(L)$ a real projective
bundle on $M$ whose Dixmier-Douady class is equal to $d(L)$ induces
a bijection
$$
\frac{\Mod_\R(L)}{\Lin_\R(M)} \to \Pro_\R(M, d(L)).
$$
\end{proposition}

We next give some examples of real bundle gerbe modules.
The first example involves Clifford algebras and spinors.
Suppose that $P$ is a principal $SO(n)$ bundle that does not have a
$Spin(n)$-structure i.e. $w_2(P)\ne 0$, for example, $P$ can be the
principal bundle of oriented frames  on a manifold $M$ with $w_2(M)\ne  
0$.
We can consider the lifting real bundle  gerbe and obtain a principal
$\Z_2$ bundle $L\to P^{[2]}$ (or equivalently a real
line bundle over $P^{[2]}$). It is natural to
call this the $Spin$ real bundle gerbe.  The
Dixmier-Douady invariant of the $Spin$ real bundle
gerbe coincides with the second Stieffel-Whitney class of $P$ in
$H^2(M, {\mathbb Z}_2)$. The pullback
$\pi^*P$ of $P$ to $P$ has a lifting to a $Spin(n)$
bundle $\widehat{\pi^*P}\to P$.  We consider the associated
bundle of spinors by taking an irreducible spin representation $V$
of $Spin(n)$ and forming the associated real vector bundle
$\cS = \widehat{\pi^*P}\times_{Spin(n)} V$ on $P$.  $\cS$ is a
bundle gerbe module for $L$, called a spinor real bundle gerbe module.
It is not hard to show that the possible  spinor bundle gerbe modules  
for
the $Spin$ bundle gerbe $L\to P^{[2]}$ are parametrised by
$H^1(M,\Z_2)$, i.e. the real line bundles  on $M$, by following
closely the proof given above in the
$Spin^{\mathbb C}$ case in \cite{MS}.

The second example involves Weyl algebras and real symplectic spinors.
Suppose that $P$ is a principal $Sp(n)$ bundle that does not have a
$Mp(n)$-structure i.e. $w_2(P)\ne 0$, for example, for example, $P$ can  
be the
principal bundle of symplectic frames  on a symplectic manifold $M$  
with $w_2(M)\ne 0$.
We can consider the lifting real bundle  gerbe and obtain a principal
$\Z_2$ bundle $L\to P^{[2]}$ (or equivalently a real
line bundle over $P^{[2]}$). It is natural to
call this the metaplectic real bundle gerbe.  The
Dixmier-Douady invariant of this metaplectic real bundle
gerbe coincides with the second Stieffel-Whitney class of $P$ in
$H^2(M, {\mathbb Z}_2)$. The pullback
$\pi^*P$ of $P$ to $P$ has a lifting to a $Mp(n)$
bundle $\widehat{\pi^*P}\to P$.  We consider the associated
bundle of symplectic spinors by taking an irreducible oscillator  
representation
on  $L^2(\RR^n)$
of $Mp(n)$ and forming the associated real Hilbert bundle
${\mathcal W} = \widehat{\pi^*P}\times_{Mp(n)} L^2(\RR^n)$ on $P$.   
${\mathcal W} $ is a
bundle gerbe module for $L$, called a metaplectic real bundle gerbe  
module.
It is not hard to show that the possible  metaplectic bundle gerbe  
modules for
the metaplectic bundle gerbe $L\to P^{[2]}$ are parametrised by
$H^1(M,\Z_2)$, i.e. the real line bundles  on $M$,

\section{$KO$-theory for real bundle gerbes}

Given a real bundle gerbe $(L, Y)$
we denote by
$KO^0_{bg}(L)$ the Grothendieck group of the
semi-group $\Mod_\R(L)$ and call this the
$KO$ group of the real bundle gerbe.
We also define $KO^j_{bg}(L) = KO^0_{bg}(p_j^*L)$
where $p_j: M\times \R^j \to M$ is the projection and
$p_j^*L$ is the pullback real bundle gerbe over $M\times \R^j $.
By Proposition  \ref{prop:module}, one has:

\begin{proposition}
               A choice of stable isomorphism from $L$ to $J$ defines a
               canonical isomorphism $KO_{bg}^j(L) \simeq KO_{bg}^j(J)$
for all $j \ge 0$.
               \end{proposition}

Notice that the group $KO_{bg}^\bullet (L)$ depends only on the
Dixmier-Douady class $d(L) \in H^2(M, \ZZ_2)$
and for any class $[H]$ in $H^2(M, \ZZ_2)$ we can define a real
bundle gerbe $L$ with
$d(L)= [H] $ and hence the groups $KO_{bg}^\bullet(L)$. When we want
to emphasise the
dependence on $[H]$ we denote
this by $KO_{bg}^\bullet(M, [H])$.

It is easy to deduce from the theory of real bundle gerbes various
        properties of this $KO$-theory:

\begin{proposition}
\label{prop:bgkprops}
Real bundle gerbe $KO$ theory satisfies the following properties:\\

(1) If $(L, Y)$ is a trivial real bundle gerbe over $M$, then
$KO_{bg}^\bullet(L) = KO^\bullet(M)$.

(2) If $(L, Y)$ is a real bundle gerbe over $M$, then
$KO_{bg}^\bullet(L)$ is a module over $KO^0(M)$.

(3)  If $(L, Y)$ and $(J, X)$ are real bundle gerbes over $M$, then
there is a homomorphism
$$
KO_{bg}^i(L) \otimes KO_{bg}^j(J) \to KO_{bg}^{i+j}(L\otimes J)
$$
where $L \otimes J$ denotes the bundle gerbe over the fibre product
of $Y$ and $X$.

(4) If $(L, Y)$ is a real bundle gerbe over $M$ and $f \colon N \to M$  
is a
continuous map,  there is a homomorphism
$$
KO_{bg}^\bullet(L) \to KO_{bg}^\bullet(f^*L)
$$
where $f^*L $ denotes the pullback bundle gerbe over $N$.

\end{proposition}

The proofs of these statements are entirely analogous for the 
corresponding Proposition 5.2 of \cite{bcmms}.  
There is another  construction that associates to any class
$[H]$ in  $H^2(M, \ZZ_2)$, groups $KO^\bullet(M, [H])$ called the
{\em twisted}
$KO$ group. Twisted
$KO$-theory shares the same properties as those in Prop.  
\ref{prop:bgkprops}.
In the next section  we discuss twisted $KO$-theory and show that
real bundle gerbe $KO$-theory and twisted $KO$-theory
coincide.

\section{Type I D-brane charges in an $H$-flux, twisted $KO$-theory
and real bundle gerbe modules}

\subsection{Twisted $KO$-theory}

We recall the definition  of twisted $KO$-theory  \cite{Ros}.
Given a class $[H] \in H^2(M, \ZZ_2)$ choose a $PO(\cH)$ bundle $Y$  
whose
class is $[H]$. It is known that that the group $Aut(\cK)$
of automorphisms of the $C^*$-algebra of real compact operators
on $\cH$ is equal to $PO(\cH)$, cf. \cite{Ros}.
We can form an associated algebra bundle,
$$
\cE = Y \times_{PO(\cH)} \cK,
$$
where $PO(\cH)$ acts on $\cK$ via the adjoint action. Then the
twisted $KO$-theory  is by definition
$$
KO^j(M, [H]) = KO_j(C_o(M, \cE))
$$
i.e. the $KO$-theory of the $C^*$-algebra of sections vanishing
at infinity of $\cE$. We recall that
\begin{equation}\label{KOdef}
KO_j(C_o(M, \cE)) = KO_0(C_o(M, \cE)\otimes C_o(\RR^j)) = KO_0(C_o(M,
\cE)\otimes Cl_j)
\end{equation}
where $Cl_j$ denotes the Clifford algebra of $\RR^j$.

Now let $\cH$ be a real infinite dimensional Hilbert space that is a
$*$-module for
the real Clifford algebra of $\mathbb R^{j-1}$, namely $Cl_{j-1}$.
Recall that when
$Cl_{j-1}$ is simple, then this representation is unique up to
equivalence, and when it
is not simple, then  it is the direct sum of two simple algebras. The
assumption made
is that each simple subalgebra is represented with infinite
multiplicity on $\cH$.
Let  $\widehat\Fred$ denote the space of all skew-adjoint Fredholm
operators on $\cH$.
For $k\ge 1$, let $\Fred_j$ denote the subspace of all skew-adjoint
Fredholm operators $\widehat\Fred$, that commute with the action of  
$Cl_{j-1}$.
Then it has been shown in \cite{AS} that $\Fred_j$ is the classifying
space for $KO^j$.
One can form the associated bundle
$$
Y(\Fred_j) = Y \times_{PO(\cH)} \Fred_j
$$
where $\Fred_j$ is acted
on by conjugation.  Let $[M, Y(\Fred_j)]$ denote the
space of all homotopy classes of sections of $Y(\Fred_j)$ then
one has, cf. \cite{Ros},
$$
KO^j(M, [H])  =  [M, Y(\Fred_j)] = [Y, \Fred_j]^{PO(\cH)},
$$
where the right hand side denotes the space
of all homotopy classes of equivariant maps with the homotopies
being by equivariant maps.

\subsection{Real bundle gerbe  $KO$-theory and twisted $KO$-theory}

Here we will
prove that real bundle gerbe $KO$-theory and twisted $KO$-theory
are the same and indicate their relationship with equivariant
$KO$-theory.

The Serre-Patterson theorem cf. \cite{Pat} says that, given
an element $[H] \in H^2(M, \ZZ_2)$, there is a principal
$PO(n)$ bundle $X\to M$, with Dixmier-Douady
invariant equal to $[H]$.
We can define an action  $SO(n)$ on $\RR^n
\otimes \cH = \cH^n$
letting $g$ act as $g \otimes 1$. This gives a
representation $\rho_n \colon SO(n) \to O(\cH^n)$
and induces  a $PO(\cH^n)$ bundle with Dixmier-Douady
     class $[H]$. As $\cH^n \simeq \cH$ and all $PO(\cH)$
bundles are determined by their Dixmier-Douady class we can
assume that this bundle is $Y$ and contains $X$ as a $SO(n)$
reduction.  Then we have
$$
(Y \times \Fred)/PO(\cH)  \cong (X \times \Fred)/PO(n),
$$
so that
$$
KO^0(M,  [H]) = [Y, \Fred]^{PO(\cH)} \cong [X, \Fred]^{PO(n)}.
$$

The lifting real bundle gerbe for $Y \to M$ pulls-back to become the  
lifting
real bundle gerbe $L$ for $X \to M$.  We will now prove that  
$KO_{bg}^0(M, L)
= KO^0(M, [H])$. Notice
that this will prove the result also for {\em any} real bundle gerbe  
with
torsion Dixmier-Douady class as we already know that real bundle gerbe  
K-theory
depends only on the Dixmier-Douady class.

In the case where there is no twist, Atiyah and Janich showed that  
$KO(M) = [M,
\Fred]$ and we
will follow Atiyah's proof indicating just what needs to be modified to
cover this equivariant case.

First we have have the following
\begin{lemma} If $W$ is  a finite dimensional subspace of $\RR^n
\otimes \cH$ there
is a finite co-dimensional subspace $V$ of $\cH$ such that $\RR^n  
\otimes V
\cap W = 0$.
\end{lemma}
\begin{proof} Let $U$ be  the image of $V$ under the map $\RR^n
\otimes \RR^n \otimes
\cH \to \cH$ where we contract the two copies of $\RR^n$ with the inner
product. Then $V \subset \RR^n \otimes U$. So take $W = U^{\perp}$.
\end{proof}

Using the compactness of $X$ and the methods in Atiyah we can show that  
if
$f \colon X \to \Fred(\RR^n \otimes \cH)$ then there is a
     subspace $V \subset \cH$, of finite co-dimension, such that
$\ker(f(x)) \cap \RR^n \otimes V = 0$. Then $\cH/V$ and $\cH/f(V)$
will be vector bundles
on $X$ and moreover they will  by acted on by $SO(n)$ in such a way as  
to make
them real bundle gerbe modules.  So we define
$$
\ind \colon [X, \Fred(\RR^n \otimes \cH ) ] _{SO(n)}\to KO_{bg}(M, L)
$$
by $\ind(f) = \cH/V -\cH/f(V)$. Again the methods of \cite{Ati} will  
show that
this index map is well-defined and a homomorphism.

As in \cite{Ati} we can identify the kernel of $\ind$ as
$ [X, O(\RR^n \otimes \cH ) ] _{SO(n)}$ and use the result of Segal  
\cite{Seg2}
which shows that $O(\RR^n \otimes \cH )$ is contractible so $\ind$ is
injective.

Finally we consider surjectivity. First we need from \cite{Seg1} the  
following
\begin{proposition}
\label{prop:sub}
If $E \to X$ is a real bundle gerbe module for $L$ then there is a
representation
$\mu\colon SO(n) \to SO(n)$ such that $E$ is a sub-real bundle gerbe  
module of
$\RR^N \otimes X$.
\end{proposition}

If $E \to X$ is a real bundle gerbe module then Proposition  
\ref{prop:sub}
enables us to find a $SO(n)$
equivariant map $\tilde f \colon X \to \Fred(\RR^N \otimes X)$ whose
index is $E$.
The action of $SO(n)$ used here on $\RR^N \otimes X$ is that induced  
from the
representation $\mu$.
To prove surjectivity of the index map it suffices to find a
map $f \colon X \to \Fred(\RR^n \otimes X)$
whose index is $E$. Then if $E-F$ is a class in $KO_{bg}(M, L)$ we can
apply a similar technique
to obtain a map whose index is $-F$ and combine these to get a map
whose index is $E-F$ and we are done.

To construct $f$ we proceed as follows.  We have a representation
$\rho_n \colon SO(n) \to \RR^n \otimes \cH$
and a representation  $\rho_N \colon SO(n) \to \RR^N \otimes \cH$.
These can be used to induce a
$PO(\RR^n \otimes \cH)$ bundle and a $PO(\RR^N \otimes \cH)$ bundle,  
both
with Dixmier-Douady class $[H]$.
So they must be isomorphic.  We need the precise form of this  
isomorphism.
Choose an isomorphism $\phi \colon \RR^n \otimes \cH \to \RR^N
\otimes \cH$. This induces
an isomorphism $O(\RR^n \otimes \cH )\to O(\RR^N \otimes \cH)$ given
by $u \mapsto
\phi u \phi^{-1}$ which we will denote by $\phi[u]$ for convenience.
There is a similar
identification $\Fred(\RR^n \otimes \cH) \to \Fred(\RR^N \otimes \cH)$.  
The two
$PU$ bundles are given by $X \times_{\rho_n} O(\RR^n \otimes \cH)$
and $X \times_{\rho_N} O(\RR^N \otimes \cH)$
and consist of cosets  $[x, u] = [xg, \rho^{-1}_n(g)u]$ and $[x, u] =
[xg, \rho^{-1}_N(g)u]$
respectively. The action of
$O(\RR^n \otimes \cH)$ is $[x, u]v = [x, uv]$ and similarly for
$O(\RR^N \otimes \cH)$. Because these
are isomorphic bundles there must be a bundle map
$$
\phi \colon X \times_{\rho_n} O(\RR^n \otimes \cH) \to
\times_{\rho_N} O(\RR^N \otimes \cH)
$$
satisfying $\phi([x,u]v) = \phi([x, u]) \phi[v]$ and hence
$\phi([x,u]) = \phi([x, 1]) \phi[u]$.
Define $\alpha \colon X \to O(\RR^N \otimes \cH)$ by requiring that
$\phi([x, 1]) =
[x, \alpha(x)]$. Then for $g \in SO(n)$, a calculation analogous to  
\cite{bcmms}
shows that
\begin{equation}\label{eq:relation}
\alpha(xg) = \rho_N(g)^{-1} \alpha(x) \phi[(\rho_n(g)].
\end{equation}

We can now define $ f \colon X \to \Fred(\RR^n \otimes X)$ by
$$
f(x)  =     \alpha(x)\phi^{-1}[ \tilde f(x) ]
$$
and it is straightforward to see that this is $SO(n)$ equivariant
by applying the equation \eqref{eq:relation}.  It is clear that
$\alpha(x)$ and $\phi$ can be used to
establish an isomorphism between $\ker(f)$
and $\ker(\tilde f)$ and hence between $\ker(f) $ and $E$.

This proves
\begin{proposition}\label{equivalence0} If $L$ is a real bundle gerbe
over $M$ with
Dixmier-Douady class $[H] \in H^2(M, \ZZ_2)$, then
$KO_{bg}^0(M, L) = KO^0(M, [H])$.
\end{proposition}

The lifting real bundle gerbe for $Y \to M$ pulls-back to become the  
lifting
real bundle gerbe for $X \to M$.  A real bundle gerbe module for
this is a bundle $E \to X$ with a $SO(n)$ action covering the
action of $PO(n)$ on $X$.  This $SO(n)$ action has to have the
property that the center $\ZZ_2 \subset SO(n)$ acts on the fibres of
$E$ by scalar multiplication.  Considered from this perspective we see  
that
we are in  the context
of equivariant $KO$ theory \cite{Seg1}. Notice that by projecting to
$PO(n)$ we can make
$SO(n)$ act on
$X$. Of course this action is
not   free, the center $\ZZ_2$ is the isotropy subgroup
at every point.
The equivariant $KO$ theory
$KO_{SO(n)}(X)$ is the $KO$ theory formed from vector bundles
on $X$ which have an $SO(n)$ action covering the action on $X$.  We
need a subset
of such bundles with a particular action.  To understand this note that
because the center $\ZZ_2 \subset SO(n)$ is the isotropy subgroup for
the $SO(n)$
action on $X$ it must act on the fibres of $E$ and hence define a
representation
of $\ZZ_2$ on $\RR^r$ if the bundle $E$ has rank $r$.  This defines an
element of
$R(\ZZ_2)$, the representation ring  of $\ZZ_2$, and to be a real
bundle gerbe module
this representation must be scalar multiplication on $\RR^r$.  In terms
of equivariant $KO$ theory we can consider the map which is
restriction to a fibre of $X \to M$ and then we have
$$
K_{SO(n)}(X) \to K_{SO(n)}(SO(n)/\ZZ_2) = R(\ZZ_2).
$$
$KO_{bg}(M, L)$ is the pre-image under this map of the representation
of $\ZZ_2$ on $\RR^n$ by scalar multiplication.

Defining $KO_{bg}^j(M, L) = KO_{bg, c}^0(M \times \RR^j, L)$,
where $KO_{bg, c}^0$ denotes compactly supported bundle gerbe
$K$-theory, we deduce from Proposition \ref{equivalence0} that

\begin{proposition}\label{equivalencej} If $L$ is a real bundle gerbe
over $M$ with
Dixmier-Douady class $[H] \in H^2(M, \ZZ_2)$, then
$KO_{bg}^j(M, L) = KO^j(M, [H])$ for all $j \ge 0$.
\end{proposition}

Using the definition in equation (\ref{KOdef}), one sees that objects in
$KO_{bg}^j(M, L) $ can also be described as differences $[E]-[F]$, where
$E, F$ are real bundle gerbe modules with an action of the real Clifford
algebra $Cl_j$.


\section{Infinite dimensional model for twisted $KO$-theory.}
\label{sec:non-torsion case}

We have seen that  twisted $KO$-theory can be defined using finite rank
real bundle gerbes. But sometimes it is useful to get an infinite  
dimensional
generalisation such as to allow real bundle gerbe modules which
are infinite dimensional real Hilbert bundles. In that case the induced
projective bundle on $M$ is a $PO(\cH)$ bundle for $\cH$
an infinite dimensional real Hilbert bundle and it is well
known that there is only one such bundle for a given
Dixmier-Douady class and hence Proposition \ref{prop:proj}
implies that

\begin{proposition}
Every real bundle gerbe admits exactly one real bundle gerbe module  
which is a
bundle of infinite dimensional Hilbert spaces  with structure group  
$O(\cH)$.
\end{proposition}

In particular if $E$ and $F$ are Hilbert real bundle gerbe modules
then $E = F$ so that the class $E=F$ in the induced
$KO$ group is zero. So the $KO$ group is zero.

In the remainder of this section
we discuss another approach to twisted cohomology
where the structure group of the real bundle gerbe module
is the group $\UK$, the subgroup of $O(\cH)$
of orthogonal operators which differ from the identity by a compact
operator (here $\mathcal{K}$ denotes the compact operators
on $\cH$).
To see how this arises notice that in
Rosenberg's definition \cite{Ros} we can replace  $\Fred$ by a
homotopy equivalent space.
For our purposes we choose $BO_{\mathcal K}\times \ZZ$.
This can be done in a $PO(\cH)$
equivariant fashion as follows. For $BO_{\mathcal K}$
we could choose the connected component of the
identity of the invertibles in the real Calkin algebra $B(\cH)/\mathcal  
K$
which is homotopy equivalent in a $PO(\cH)$ equivariant way
to the real Fredholms of index zero under the quotient map
$\pi: B(\cH)\to B(\cH)/\mathcal K$. Note however
that the identity component of the invertibles in the Calkin
algebra is just $GL(\cH)/GL_\mathcal K$ where $GL(\cH)$
denotes the invertible operators on $\cH$ and $GL_\mathcal K$
are the invertibles differing from the identity by a compact.
Thus we can take $BO_{\mathcal K}$ to be $GL(\cH)/GL_{\mathcal K}$
and this choice is $PO(\cH)$ equivariant. We could equally well take
$B\UK = O(\cH)/\UK$.

As $O(\cH)$ acts on $O_{\mathcal K}$ by conjugation there is
a semi-direct product
$$
O_{\mathcal K} \to O_{\mathcal K} \rtimes PO(\cH) \to PO(\cH).\
$$
Note that this means that any $ O_{\mathcal K} \rtimes PO(\cH)$
bundle over $M$ induces a $PO(\cH)$ bundle and hence a class
in $H^2(M, \ZZ_2)$.   If $R \to Y$ is a $O_{\mathcal K}$ bundle
we call it $PO(\cH)$ covariant if there is an action
of $PO(\cH)$ on the right of $R$ covering the action on $Y$ such that
$(rg)[u] = r[u] u^{-1}gu$ for any $r \in R$,
$[u] \in PO(\cH)$ and $g \in O_{\mathcal K}$. Here $[u]$ is the
projective class of some $u \in O(\cH)$.

Because $B\UK$ is homotopy equivalent to only the connected component
of index $0$ of $\Fred$ it is convenient to work with reduced twisted
$KO$ theory, $\redK O(M, [H])$, defined by
$$
\redK O(M, [H]) = [M, Y(B\UK)].
$$

We have
\begin{proposition}
Given a $PO(\cH)$ bundle $Y \to M$ with Dixmier-Douady class $[H] \in
H^2(M, \ZZ_2)$ the
following are equivalent, with a proof analogous to that given in  
\cite{bcmms}.
\begin{enumerate}
\item $\redK O(M, [H])$
\item space of homotopy classes of sections of $Y \times_{PO(\cH)}
BO_{\mathcal K}$
\item space of homotopy classes of  $PO(\cH)$ equivariant maps from $Y$  
to
$BO_{\mathcal K}$
\item space of isomorphism classes of $PO(\cH)$ covariant $O_{\mathcal  
K}$
bundles on $Y$, and
\item space of isomorphism classes of $O_{\mathcal K} \rtimes PO(\cH)$
bundles on $M$ whose projection to a $PO(\cH)$ bundle has class $[H]$.
\end{enumerate}
\end{proposition}

\begin{note} Notice that if we worked with $\Fred$ instead
of $B\UK$ then it has connected components $\Fred_n$ consisting
of operators of index $n$.  We can then consider sections
of $Y \times_{PO(\cH)} \Fred_n$ for every $n$, not just zero.
Such a section will pull back a $KO$ class and if we take
the determinant of this $KO$ class we will obtain a line bundle
on $Y$ on which the gerbe $L^n$ acts.  Hence we will have
$n [H] = n d(L) = 0$ and so
we deduce the result noted in \cite{Ati2} that if $[H] $ is
not torsion then there are no sections of $Y \times_{PO(\cH)} \Fred_n$
except when $n = 0$ so $\redK O(M, [H]) = KO(M, [H])$.
\end{note}

\subsection{$O_{\mathcal K}$ real bundle gerbe modules}

Given a $PO(\cH)$ covariant $O_{\mathcal K}$ bundle $R$ over $Y$ we can
define the associated bundle
\begin{equation}
\label{eq:associated}
E = R \times_{O_{\mathcal K}} \cH \to Y.
\end{equation}
We claim that this is a real bundle gerbe module for the
lifting real bundle gerbe $P$.
Let $[r, v] \in E_{y_1}$ be a $O_{\mathcal K}$ equivalence class
where $r \in R_{y_1}$, the
fibre of $R$ over $y_1 \in Y$  and $v \in \cH$.
Let $u \in L_{y_1y_2}$ be an element of the lifting bundle
gerbe. Then, by definition, $u \in O(\cH)$ and $y_1[u] = y_2$.  We
define the action of $u$ by $[r, v]u = [r[u], u^{-1}v]$.  It is
straightforward to check that this is well defined.  Hence we have
associated to any $PO(\cH)$ covariant $O_{\mathcal K}$ bundle $R$ on $Y$
a module for the lifting real bundle gerbe.

The inverse construction is also possible if the real bundle gerbe
module is a $\UK$ real bundle gerbe module which we now define.
Let $E\to Y$ be a  Hilbert bundle with structure
group $\UK$.  We recall what it means for a Hilbert bundle
to have structure group $\UK$.  To any Hilbert bundle  there is   
associated a
$O(\cH)$ bundle
$O(E)$ whose fibre, $O(E)_y$, at $y$ is all unitary isomorphisms
$f \colon \cH \to E_y$. If
$u \in O(\cH)$ it acts on $O(\cH)_y$ by $fu = f \circ u$ and hence  
$O(E)$ is
a principal $O(\cH)$ bundle.  For $E$ to have structure
group $\UK$ means that we have a reduction of $O(E)$ to an $\UK$ bundle
$R \subset O(E)$.
Each  $R_y \subset O(E)_y$ is an orbit under $\UK$, that
is $R$ is a principal $\UK$ bundle.

For $E$ to be an $\UK$ real bundle gerbe module we need to define
an action of the real bundle gerbe on it.  By comparing with the action  
on the
bundle $E$ defined in \eqref{eq:associated} we see that we need to make  
the
following  definition.  If
$u \in O(\cH)$ such that $y_{1}[u] = y_{2}$  then $u \in L_{(y_1, y_2)}$
where $L \to Y^{[2]}$ is the lifting real bundle gerbe so if $f \in
R_{y_1}$ then
$ufu^{-1} \in O(E)_{y_2}$.  We require that $ufu^{-1} \in R_{y_2}$. So
a lifting real bundle gerbe module which is an $\UK$ Hilbert bundle and
satisfies this condition we call an $\UK$ real bundle gerbe module.  By
construction
we have that the associated $R$ is an $\UK$ bundle over $Y$ on which
$PO(\cH)$ acts.  Let us denote by $\Mod_{\UK}(M, [H])$
the semi-group of all $\UK$ real bundle gerbe modules for the
lifting real bundle gerbe of the $PO(\cH)$ bundle with Dixmier-Douady class  
$[H] \in H^2(M;\ZZ_2)$ . As
any two $PO(\cH)$ bundles with the same Dixmier-Douady class are isomorphic we  
see
that $\Mod_{\UK}(M, [H])$ depends only on $[H]$.

We have now proved
\begin{proposition}
\label{prop:U_K bg modules equals KO(M;H)}
If $(L, Y)$ is the lifting real bundle gerbe for a $PO(\cH)$
bundle with Dixmier-Douady class $[H]$
$$
\redK O(M, [H]) =  \Mod_{\UK}(M, [H]) .
$$
\end{proposition}

If $L_1$ and $L_2$ are two $PO(\cH)$ covariant $O_{\mathcal K}$ bundles
on $Y$ note that
$L_1 \times L_2$ is an $O_{\mathcal K} \times O_{\mathcal K}$ bundle.
Choose an
isomorphism $\cH \times \cH \to \cH$ which induces an isomorphism
$O_{\mathcal K} \times O_{\mathcal K} \to O_{\mathcal K}$ and hence  
defines a
new $PO(\cH)$ covariant bundle $L_1 \otimes L_2$.  It is straightfoward
to check that
$$
(L_1 \otimes L_2 )(\cH) \simeq L_1(\cH) \times L_2(\cH).
$$

This makes $\Mod_{\UK}(M, [H])$ a semi-group and the map
$\redK O(M, [H]) =  \Mod_{\UK}(M, [H])$ is a semi-group isomorphism.
Note that with our definition $B\UK$ is a group. Moreover
the space of all equivariant maps $Y \to B\UK$ is a group as well.
To see this notice that if
     $f$ and $g$ are equivariant maps and we multiply
pointwise then for  $y \in Y$ and $[u] \in PO(\cH)$ we have
\begin{align*}
(fg)(y[u]) &= f(y[u]) g(y[u]) \\
              &= (u^{-1} f(y) u) (u^{-1} g(y) u) \\
              &= u^{-1}(fg)(y) u
\end{align*}
and if $f^{-1}$ is the pointwise inverse then  $f^{-1}(y[u]) =
\left( u^{-1} f(y) u \right)^{-1} = u^{-1} f^{-1}(y) u$.
This induces a group structure on $\Mod_{\UK}(M, [H])$. We have already
noted that it is a semi-group but this implies more, for every $\UK$  
bundle
gerbe module $E$ there is an $\UK$ real bundle gerbe module
$E^{-1}$ such that $E \oplus E^{-1}$
is the trivial $\UK$ real bundle gerbe. Hence we have

\begin{proposition} If $(L, Y)$ is the lifting real bundle gerbe for
a $PO(\cH)$
bundle with Dixmier-Douady class $[H]$ then
$$
      K_{\UK}(M, [H]) = \Mod_{\UK}(M, [H])  = \redK O(M, [H])
$$
\end{proposition}

\begin{note}
\label{note:replace U_K with U_1}
(1) The group $\UK$ used here could be replaced by any
other group to which it is homotopy equivalent by a homotopy
equivalence preserving the $PO(\cH)$ action.  In particular we
could consider $O_1$, the subgroup of $O(\cH)$ consisting
of orthogonal operators which differ from the identity by a
trace class operator.
In Section 9 we show that the computation in Section 6 of real bundle  
gerbe
characteristic classes generalizes, with some modifications, to
$O_1$ real bundle gerbe modules.

\end{note}

\subsection{Local description of $\UK$ real bundle gerbe modules}
Let $\{ U_i \}_{i\in I}$ be a good cover of $M$ and let $U_{ij\dots
k} = U_{i} \cap U_{j} \cap \dots \cap U_{k}$.
The trivial bundle has a sections $s_i$ which are related by
$$
s_i = s_j [u_{ji}]
$$
where $[u_{ji}] \colon U_{ij} \to PO(\cH)$ for some $u_{ji} \colon
U_{ij}\to O(\cH)$ where
$u_{ij}u_{jk}u_{ki} = g_{ijk}1$ where $1$ is the
identity operator and the $g_{ijk}$ are non-zero scalars.

Over each of the $s_i(U_i)$  are sections $\sigma_i $ of the $\UK$
bundle $R$. We can compare
$\sigma_i$ and $\sigma_j[u_{ji}]$ so that
$$
\sigma_i = \sigma_j [u_{ji}] g_{ji}
$$
where $g_{ij} \colon U_{ij} \to \UK$. These satisfy
\begin{equation}
\label{eq:twisted}
g_{ki} = ([u_{ji}^{-1}] g_{kj} [u_{ji}]) g_{ji}.
\end{equation}

If $Y_i = \pi^{-1}(U_i)$ one can define a section of $R$ over all of  
$Y_i$ by
$\hat\sigma_i ( s_i [u]) = \sigma_i[u]$.  The transition functions for  
these
are $\hat g_{ij}$ where $\hat g_{ij} (s_j[u]) = [u^{-1}] g_{ji} [u]$  
and the
identity \eqref{eq:twisted}  is equivalent to $\hat g_{ki} = \hat
g_{kj} \hat g_{ji}$.


\section{Continuous trace real $C^*$-algebras}

There is a well-known construction of a continuous
trace real $C^*$ algebra from
a groupoid \cite{Ren}, \cite{Ros}.  This can be used to construct a
$C^*$ algebra
from real bundle gerbes as follows.  If the fibres of
         $Y \to M$ have an appropriate Haar measure and we can
define a product on two sections $f, g \colon  Y^{[2]} \to P$ by
$$
(fg)(y_1, y_2) = \int f(y_1, y)g(y, y_2) dy
$$
where in the integrand we use the real bundle gerbe product so that
$f(y_1, y) g(y, y_2) \in L_{(y_1, y_2)}$.
Closing this space of sections in the operator norm topology
gives a continuous trace real $C^*$ algebra with spectrum $M$
and Dixmier-Douady class the Dixmier-Douady class of $(L, Y)$.
Some constructions in the theory of $C^*$ algebras become easy from this
perspective. For example if $A$ is continuous trace real $C^*$-algebra
with spectrum $X$ and $f \colon Y \to X$
is a continuous map, then there is continuous trace real
$C^*$-algebra $f^{-1}(A)$ with spectrum $Y$.
This is just the pullback of real bundle gerbes.  Then
the $KO_{bg}$-theory is just the $KO$-theory of the
associated continuous trace real $C^*$-algebra.
This works as long as we take finite dimensional real bundle gerbes,
which we know represent all classes in $H^2(M, \Z_2)$.

\section{Type I D-brane charges $\Rightarrow$ Type II D-brane charges}

Here we discuss the natural complexification map from twisted  
$KO$-theory
to twisted $K$-theory, which can be used to define the twisted Chern  
character
homomorphism in twisted $KO$-theory. This homomorphism is analysed in
some detail, together with examples.

\subsection{The complexification homomorphism}

We use the well known and fundamental fact that the complexification
of a real vector vector bundle
is a complex vector bundle that is isomorphic to its own conjugate
vector bundle., cf. \cite{Ati}.
We begin with a real bundle gerbe $(L, Y)$ over $M$, where
$\pi : Y \to M$ is a locally trivial fibre bundle and $L\to Y^{[2]}$ is  
a real
line bundle satisfying equation \eqref{gerbe}. The complexification
of the real bundle $(L,Y)$
determines the bundle gerbe $(L\otimes \C, Y)$ over $M$ which is
automatically isomorphic
to $(\overline{L\otimes \C}, Y)$. If the Dixmier-Douady invariant
of $(L, Y)$ is $d(L) \in H^2(M, \Z_2)$, then it is easy to see that
the Dixmier-Douady invariant
of $(L\otimes \C, Y)$ is $d(L\otimes \C) \in H^3(M, \Z)$, where
$d(L\otimes \C) =
\beta(d(L))$ and $\beta :  H^2(M, \Z_2) \to H^3(M, \Z)$ is the
Bockstein homomorphism.
Observe also that $d(L\otimes \C) = d((\overline{L\otimes \C}) = -
d({L\otimes \C})$.
The complexification map is compatible with stable isomorphism of
real and complex
bundle gerbes, and therefore defines a homomorphism from stable  
equivalence
classes of real bundle gerbes and stable equivalence
classes of bundle gerbes.

Now let $(L, Y)$ be a real bundle gerbe
over a manifold $M$ and let $E \to Y$
be a finite dimensional real bundle gerbe module.
Then it is straightforward to see that the complexification
$E\otimes \C \to Y$ is a bundle gerbe module for the
bundle gerbe $(L\otimes \C, Y)$ over $M$. This map is
compatible with real and complex bundle gerbe stable
equivalences, and isomorphism of real and complex
bundle gerbe modules, therefore defining a well defined
homomorphism,
\begin{equation}\label{complexi}
\otimes\C : KO^0(M, d(L)) \rightarrow K^0(M, d(L\otimes \C))
\end{equation}
where we recall that $d(L\otimes \C) = \beta(d(L))$ as above.

We can use this homomorphism to define the twisted Chern character,
\begin{equation}\label{realchern}
ch_{d(L)} :  KO^0(M, d(L)) \to H^{even}(M, \R)
\end{equation}
by $ch_{d(L)} (E) := ch_{d(L\otimes \C)}(E\otimes \C)$, where we have
used the twisted chern character as defined in \cite{bcmms}.
Explicitly, it is determined by the property $$\pi^*(ch_{d(L)} (E)) =
e^{c_1(L\otimes\C) } ch(E\otimes \C). $$
It is a
homomorphism with the usual properties, namely,\\
1) $ch_{d(L)}$ is natural with respect
to pullbacks, \\
2) $ch_{d(L)}$ respects the ${KO}^0(M)$-module
structure of ${KO}^0(M, d(L))$,  \\
3) $ch_{d(L)}$ reduces
to the ordinary Chern character in the untwisted case
when $d(L) = 0$.  \\

Since $E\otimes \C \to Y$ is isomorphic to $\overline{E\otimes \C} \to  
Y$
for every real bundle gerbe module $E\to Y$, we see that
$$
   \pi^*(ch_{d(L)} (E)) = e^{c_1(L\otimes\C) } ch(E\otimes \C ) =
e^{c_1(\overline{L\otimes\C})} ch(\overline{E\otimes \C})
$$
from which one deduces that all of the components
of the twisted Chern character of degree $4j+2$ vanish. Therefore,
\begin{equation}\label{realchern}
ch_{d(L)} :  KO^0(M, d(L)) \to H^{4\bullet}(M, \R)
\end{equation}
and is an isomorphism over the reals.

\subsection{Examples for bundle gerbe $K$-theory: the non-torsion case}
Here we give generators in non-torsion twisted complex   
$K$-theory, and
also examples in twisted $KO$-theory.
These are mainly for the case when the twisting class is decomposable.
Related bundle gerbe constructions are in \cite{Jo}.

\subsubsection{Twisted self-adjoint Fredholm familes and twisted $K^1$}
We begin by defining a canonical homomorphism
\begin{equation}\label{det}
\det : K^1(M, [H]) \to H^1(M, \Z).
\end{equation}
Recall that $ K^1(M, [H])$ is represented by equivariant 
continuous maps $G: P \to  
U_{\tr}$,
where  $U_{\tr}$ denotes the unitaries that differ from the  identity
by a compact operator and where $P\to M$ is a principal 
$PU$ bundle with Dixmier-Douady class equal to $[H]$.
Such a map has a determinant, which defines a continuous map
$\det(G) : M \to U(1)$ since the determinant is invariant under the  
conjugation
action of the unitary group. This defines the homomorphism in  
(\ref{det}).

Take $L$ to be a line bundle on $S^2$ such that the first Chern class
   $c_1(L) = b \in H^2(S^2, \Z)$. Let $a \in H^1(S^1, \Z)$ be a non-zero  
element;
then $a$ can be thought of
as a character $\chi_a : \pi_1(S^1) \to \Z$. Let  $\Z\to \R_a \to
S^1$ be the principal
$\Z$-bundle corresponding to this
character that is $\R_a =  (\R \times_a \Z)$.
Consider the principal $\Z$-bundle $Y = \R_a \times  S^2 \to S^1 \times  
S^2$.
$$
Y^{[2]} = \R_a \times  \Z \times  S^2
$$
and the bundle gerbe over $(t, m, x)$ is $L_x^m$, where $L^m =  
L^{\otimes m}$.
The groupoid  multiplication is $(t, m , x) (t, n , x) \to (t, m+n, x)$
and lifts to   $L_x^m \otimes L_x^n \to L_x^{m+n}$.

Define $\cE \to Y$ a Hilbert bundle by $\cE_{(t,x)} =
\bigoplus_{n=-\infty}^\infty L^n_x$,
where the right hand side denotes the Hilbert direct sum,
defined using a hermitian inner product on
$L$ and hence on $L^n$.
The bundle gerbe action on $\cE$ is  the obvious shift operation and
bundle gerbe multiplication.
Now define  $F(t, x) : \cE_{(t,x)} \to \cE_{(t,x)}$ by $F(t, x)(v_n)
= f(t-n)v_n$ if
$v_n \in L_x^n$, where $f: \R_a \to \R_a$ is a bounded uniformly
continuous function.
This ensures that we get a twisted norm continuous family of  
self-adjoint
Fredholm operators.
We need to know this commutes with the bundle gerbe action.
Let $v_m \in L_x^m$, then the bundle gerbe action
is
$v_m . v_n = v_m v_n$   where this latter multiplication is the
tensor product $L^m\otimes L^n \to L^{m+n}$.
Hence we need to show    $v_m F(t, x) v_n = F(t+m, x) v_mv_n$.
But the LHS is
$$
v_m f(t-n) v_n = f(t-n) v_mv_n
$$
and the RHS is
$$
f(t+m - (m+n)) v_mv_n = f(t-n)v_m v_n
$$

Notice that if $t$ is not an integer $F(t, x)$ is an isomorphism
and if $t$ is an integer $F(t, x)$ has 1 dimensional kernel and
1 dimensional cokernel. That is, $F(t, x)$ is a Fredholm operator
for all $(t, x) \in \R_a \times S^2$, and since it commutes with the
bundle gerbe action, it determines an element in the odd degree
bundle gerbe $K$-theory $K^1(L, Y)$. We will compute the
degree one component of the odd dimensional
Chern character of this element and show that it is not trivial.

We digress to discuss twisted cohomology.
Suppose $H$ is a closed differential $3$-form on
a compact manifold $M$.
We can use $H$ to construct a differential $\delta_H$ on the
algebra $\Omega^\bullet(M)$ of differential forms
on $M$ by setting $\delta_H(\omega) = d\omega -H\omega$
for $\omega \in \Omega^\bullet(M)$.  It is easy to check
that indeed $\delta_H^2 = 0$.  If $\omega \in \Omega^{\text{odd}}(M)$
so that $\omega = \omega_1 + \omega_3 +\omega_5 +\ldots$,
where $\omega_j \in \Omega^j(M)$,
then $\omega$ is in the kernel of $\delta_H$ if  the
following set of equations are satisfied,
\begin{eqnarray*}
d\omega_1 &=& 0 \\
d\omega_3 &=& H\omega_1 \\
d\omega_5 &=& H\omega_3 \\
&  \vdots &
\end{eqnarray*}
Thus the degree 1-component of $\omega$,
$\omega_1$, is always a closed form.  In particular, the degree
1-component of the twisted Chern character is always a closed
1-form.

The degree one component of the Chern character of the
twisted self-adjoint Fredholm family is $c_1^{odd}(F) = exp(2\pi i
\eta(F))^*\Theta$,
where $\Theta$ is the standard generator of the first cohomology
of $U(1)$.
If we choose $f$ to be the sign function, then $exp(2\pi i
\eta(F))^*\Theta$ can be
calculated using a result by Gilkey, (\cite{Gilkey} Section 1.10)  
to be $2 pr_1^*\Theta \ne
0$,  where $pr_1: S^1 \times S^2
\to S^1$ is the projection onto the first factor.
In general, if the function $f$ is chosen to take on both positive
and negative values on the set $A = \{t-n: n\in \mathbb N \}$  
infinitely often
then the construction above is likely to also yield a non-trivial
class in twisted
$K$-theory.
Therefore $(\cE, F)$ defines a non-trivial, in fact non-torsion, class
in $K^1(S^1 \times S^2, \xi)$, where $\xi = a \cup b \in H^3(S^1
\times S^2, \Z)$.
This example is motivated by the Schr\"odinger representation
of the Heisenberg group given by a $U(1)$ extension of $S^1\times\Z$.
This construction extends without significant change to the general
case given by
a product of two manifolds.

\subsubsection{Decomposable case: generator in twisted $K^1$}

Here we extend the construction to the case when the cohomology
class on the manifold $M$ is the product of a 1-class and a 2-class.
Assume that $M$ is endowed  with a good cover. Locally a
line bundle has transition functions $g_{ab}$ and the 1-class
comes from a $\Z$ cocycle $n_{ab}$ with $n_{ab} = s_a - s_b$ where the
$s_a$ are real valued functions.  One verifies that
$$
w_{abc} = g_{ab}^{n_{ab}} g_{bc}^{n_{bc}} g_{ca}^{n_{ca}}
$$
is the class in $H^2(M, \underline{U(1)})$ which is the DD class of the  
bundle
gerbe we want and is decomposable being the cup product of the 1 class  
and
the 2-class defined above.  This example also features in
Brylinski's work \cite{Bry}.

Let $\cE$ be the Hilbert bundle $\bigoplus L^j$ with fibre $\cH$
   and $T$ a shift operator relative
some orthnormal basis $e_n$ i.e. $T(e_n) = e_{n+1}$.
Define   $G_{ab} = (g_{ab} T)^{n_{ab}}$.
This is a unitary and  $$G_{ab}G_{bc}G_{ca} = w_{abc} .1$$
So the collection of $G_{ab}$ defines a $PU(\cH)$ bundle $P$ over $M$.
Now define $F_a$ a local Fredholm map by
$$
F_a (e_n) = f(s_a - n) e_n
$$
where $f: \R\to \R$ is a bounded measurable function.
A computation as in  the previous example shows that
$$
F_a = G^{-1}_{ab} F_b G_{ab},
$$
so this is a well-defined Fredholm map
commuting with the bundle gerbe actions.

To calculate the degree one component
of the odd Chern class, one proceeds as follows.
For simplicity, we choose the function $f$ to be the sign
function.
The universal case is given by the product construction given
in section 3.1, but applied to $S^1 \times BS^1$. In that case,
the degree 1 component of the odd Chern character is given by
$2  pr_1^*\Theta \ne 0$. Now the decomposable class is given
by a map $\Lambda : M \to S^1 \times BS^1$. By the naturality
property of the Chern character, we see that the degree 1
component of the odd Chern character of $(\cE, F)$ is given by
$2 \Lambda^* pr_1^*\Theta = 2 [n_{ab}] \ne 0 $.

\subsubsection{Twisted Fredholm families and twisted $K^0$}\label{twistK0}

We begin by defining a canonical homomorphism
\begin{equation}\label{det0}
\det : K^0(M, P) \to H^2(M, \Z).
\end{equation}
Recall that $ K^0(M, P)$ is represented by differences $E - F$,
where $E$, $F$ are Hilbert bundles over $P$, with given reduction
of structure group to $ U_{\tr}$ and with an action of the lifting  
bundle
gerbe associated to the principal $PU$-bundle $P$. We first observe that
there are canonical determinant line bundles on $P$ determined by $E$,  
$F$ and
defined as $\det(E) = P_E \times_{U_{\tr}} U(1)$ and $\det(F) = P_F
\times_{U_{\tr}} U(1)$.
Then $\det(E)  \otimes \det(F)^*$ is a  line bundle on $P$ that is
invariant under the
action of $PU$, so it descends to a line bundle on $M$. This defines
the homomorphism in (\ref{det0}).

Let $L\to X$ denote a line bundle whose first Chern class is equal
to $b \in H^2(X, \Z)$.
Define the Hilbert direct sum of line bundles,
$$
E_m = \bigoplus_{n \geq m} L^n,
$$
which determines a Hilbert bundle over $X$. Then we have the exact  
sequence
$$
    E_2 \to E_1 \to L
$$
where the map $F: E_1 \to E_2$ is inclusion. $F$ is a Fredholm
map which realizes a non-trivial element in $K^0(X)$ whenever
$b \ne 0$. In particular, it realizes a generator in $K$-theory over
$X$ when $c_1(L)$ is a generator of $H^2(X, \Z)$: this is
since ${\rm ker }(F) = 1$ and ${\rm coker} (F) = L$ and so
${\rm index}(F) = [1] - [L] \in K^0(X)$ is a nontrivial generator.

We can cover $S^1$ by two charts $U_a, U_b$.
On the  overlaps, the transition function $g_{ab} = b^{n_{ab}}
: X \to PU$, where $n_{ab}$ is a C\v ech representative
of the class $p_1^*a$ where $a \in H^1(S^1, \Z)$.
All of this can be pulled back to $U_a \times X$
and $U_b  \times X$ by the
projection map onto $X$ --- we call the pulled back
bundles $E_j^a, E_j^b$ respectively. Choose trivializations
$\phi_{a, j} : E_j \to U_a \times X \times \cH$ and
$\phi_{b, j} : E_{n_{ab}+j} \to U_b \times X \times \cH$. Then we have
$g_{ab} \phi_{a, j} = \phi_{b, j}$ for $j=1,2$.

If we denote by $F_a: E_1^a \to E_2^a$ and $F_b: E_{n_{ab}+1}^b \to
E_{n_{ab}+2}^b$
the inclusion maps, which are Fredholm as discussed earlier, then we  
have
$g_{ab}\big[ \phi_{a, 1} \circ F_a \circ \phi_{a, 2} \big]
= g_{ab}\phi_{a, 1}  \circ F_a  \circ \phi_{a, 2} g_{ab}
=  \phi_{b, 1} \circ  F_b \circ \phi_{b, 2} .
$ That is, $\{F_a, F_b\}$ defines a section of twisted Fredholm
operators, where the twist is the principal $PU$ bundle defined
by the transition function $b^{n_{ab}}$, and whose Dixmier-Douady
class is $a\cup b \in H^3(S^1 \times X, \Z)$.

Here is an argument which shows that the element of $K^0(S^1\times
X;a\cup b)$ defined by the section of the bundle of twisted
Fredholm operators above is non-zero.  We have an inclusion
$i\colon U_a\times X\to S^1\times X$.  The class $a\cup b$ vanishes
when pulled back to $U_a\times X$ and hence the twisted
$K$-group $K^0(U_a\times X;i^*(a\cup b)) = K^0*(U_a\times X)$.  The
image of the section determined by the pair $\{F_a,F_b\}$ under the
map $i^*\colon K^0(S^1\times X;a\cup b)\to K^0(U_a\times X)$ is
the element of $K^0(U_a\times X)$ determined by the map
$F_a\colon U_a\times X\to \Fred$.  This has non-zero index equal to
$[1] - [L]$.

\subsubsection{Decomposable case: generator in twisted $K^0$}

Here we extend the construction in the previous subsection
to the case when the cohomology
class on the manifold $M$ is the product of an integer
1-class $a$ and an integer 2-class $b$.

As before, we notice that the universal space for decomposable classes
is  $S^1 \times BS^1$, for which the product space construction of
section \ref{twistK0} applies. Since $a$ determines a map $f_a: M \to  
S^1$
and $b$ determines a map $f_b : M \to BS^1$, the Cartesian product
$f_a \times f_b : M \to S^1 \times BS^1$ is a continuous map, which
can be used to pullback the section of twisted Fredholm
operators on $S^1 \times BS^1$ to $M$.


\subsection{Remarks on the spectral sequence relevant to calculations} 

We will now briefly discuss some calculations of twisted $KO$-groups of
compact surfaces, in terms of the untwisted $KO$-groups. The first 
non-trivial
differential in the spectral sequence as described in \cite{Ros} is
$d_2 = H + Sq^2$, where $Sq^2$ denotes the second Steenrod square,
which vanishes for compact surfaces.
Therefore $d_2 = H$. Again for compact surfaces, there are no other
non-trivial differentials. It follows that $KO^i(\Sigma, H) \cong KO^i 
(\Sigma)$
whenever $i \ne -2, -3$, since $KO^{-i}(pt)$ is equal to $\ZZ, \ZZ_2, 
\ZZ_2, 0,\ZZ,
0,0,0$ when $i$ ranges from $0$ to $7$.
Now let $\Sigma$ be a compact surface and $[H] \in H^2(\Sigma, \ZZ_2) 
\cong \ZZ_2$
be nontrivial.  Then following the spectral sequence, one obtains
$KO^{-2} (\Sigma, H) \cong KO^{-2}(\Sigma)/\ZZ_2$ and
also $KO^{-3} (\Sigma, H) \cong KO^{-3}(\Sigma)/\ZZ_2$.

\section{Concluding remarks}

In  this paper we have developed the foundations of real bundle gerbes  
and
the geometric inteterpretation of twisted $KO$-theory as the $KO$-theory
of real bundle gerbe modules, as it may prove useful in
the geometric understanding of D-brane charges and fields in Type I
string theory. We have also studied the fundamental properties
of twisted $KO$-theory including the Chern character as well as its
relation to twisted $K$-theory etc.

In some work in progress, we study the
real index bundle gerbe, called the Pfaffian gerbe,
that is associated to a family of skew adjoint real
Dirac type operators, together with a natural metric on it as well as
a compatible connection and curving
for it. We also relate it to the (complex) index bundle gerbe
under the complexification homomorphism.
This serves as one of the motivations for this paper.

An interesting open question that remains unanswered is the following.
A missing Chern-Weil or geometric interpretation
of the Dixmier-Douady invariant of real bundle gerbes
when the group ${\rm Ext}(H_1(M, \ZZ), \ZZ_2)$ does not vanish:
this can happen whenever the torsion subgroup of $H_1(M, \ZZ),$
has 2-torsion elements as discussed in section \ref{gerbe hol}, equation
\eqref{uct}.
  

\end{document}